# Nutation spin waves in ferromagnets

Sergei V. Titov,[1] William J. Dowling,[2] Yuri P. Kalmykov[3] and

Mikhail Cherkasskii[4] [*]

[1] *Kotel'nikov Institute of Radio Engineering and Electronics of the Russian Academy of Sciences, Vvedenskii Square 1, Fryazino, Moscow Region, 141190, Russia*

[2] *Department of Electronic and Electrical Engineering, Trinity College Dublin, Dublin 2, Ireland*

[3] *Laboratoire de Modélisation Pluridisciplinaire et Applications, Université de Perpignan Via Domitia, F-66860, Perpignan, France*

[4] *Faculty of Physics, University of Duisburg-Essen, Duisburg, 47057, Germany*

**Abstract**

Magnetization dynamics and spin waves in ferromagnets are investigated using the inertial Landau-Lifshitz-Gilbert equation. Taking inertial magnetization dynamics into account, dispersion relations describing the propagation of nutation spin waves in an arbitrary direction relative to the applied magnetic field are derived via Maxwell's equations. It is found that the inertia of magnetization causes the hybridization of electromagnetic waves and nutation spin waves in ferromagnets, hybrid nutation spin waves emerge, and the redshift of frequencies of precession spin waves is initiated, which transforms to precession-nutation spin waves. These effects depend sharply on the direction of wave propagation relative to the applied magnetic field. Moreover, the waves propagating parallel to the applied field are circularly polarized, while the waves propagating perpendicular to that field are elliptically polarized. The characteristics of these spin nutation waves are also analyzed.

[*] Corresponding author: macherkasskii@hotmail.com

# I. INTRODUCTION

For decades, nutation of magnetization was neglected and considered as a paltry effect. Indeed, the majority of resonance and wave phenomena in magnetic materials was well described under the assumption that precession of magnetization dominates over nutation. However, the recent experimental evidence of nutation [1] at terahertz frequency necessitates a revision of the established point of view. The observation of nutation confirmed the previous theoretical predictions that the analogy between a spinning top and magnetization is more than a pedagogical concept [2] and nutation must be accounted for in the theory of magnetization dynamics. As in the case of a rigid body, the origin of nutation of magnetization is due to inertia of magnetization [3] leading to noncollinearity between angular momentum and magnetization [2]. Beyond the mechanical approach, other models demonstrated inertial behavior of magnetization, among which are the relativistic quantum framework [4], the torque-torque correlation model [5], mesoscopic theory [6], and the breathing Fermi surface model [7]. However, the earlier theoretical models afford understanding of experimental results, but by merging nutation into different phenomena.

The investigation of inertia indicated that it manifests as a nutation resonance in the terahertz frequency range [8-12] and causes nutation spin waves (NSWs) [13-15] as well as inducing a frequency shift of the ordinary (precession) spin waves [13]. The latter can be considered as a sequence of phase shifts between precessing magnetic moments coupled by exchange and dipole interactions; these interactions provide potential energy. The inclusion of inertia brings kinetic energy and an additional degree of freedom in spin systems, and that eventually leads to the emergence of NSW. These waves were introduced as collective excitations in Heisenberg spin chains possessing one-particle behavior – nutation acquiring mass via the Brout-Englert-Higgs mechanism [14]. The surface nature of nutation waves and their relatively low group velocity were predicted in ferromagnetic films [13]. More recently, in the investigation of nutation waves in nanostructures with uniaxial anisotropy, it was discovered that they can be excited magnetoelastically [15]; in this study a special case of longitudinal propagation was implicitly considered. However, these studies are clearly insufficient to fully characterize wave phenomena caused by nutation.

In contrast, the linear phenomena of propagation precession spin waves have been comprehensively investigated in ferromagnetic films [16,17], layered magnetic structures [18,19], and magnonic crystals [20,21]. It was found that these waves exhibit either dispersive or nondispersive propagation, isotropic or anisotropic propagation, and nonreciprocity and inhomogeneous medium effects [16,22-25]. In addition, precession spin waves exhibit an abundance of nonlinear phenomena [26-30], whereas linear and nonlinear effects of terahertz NSWs have not been studied sufficiently. This study is significant not only from a fundamental point of view, but also from an applied perspective, since nutation manifests itself at terahertz frequencies that are promising for technological applications [31].

Here we investigate dispersion relations of nutation spin waves propagating in an arbitrary direction, taking into account exchange coupling and electromagnetic properties of ferromagnets over the entire wave number range. We assume that the dielectric permittivity is a scalar quantity, while the conductivity and the shape anisotropy are negligible. These assumptions allow us to find an additional dispersion branch of NSWs, show the hybridization between electromagnetic waves and NSWs and determine the nutational-induced spectral shift of precession spin waves. Moreover, we demonstrate that the magnetization trajectories are transformed from circular to starshaped.

In order to include magnetization inertia, the Landau-Lifshitz-Gilbert (LLG) equation was generalized by including an additional term that is the second-order time derivative of the magnetization with the appropriate coefficient. The resulting equation is given by

$$\frac{d\mathbf{M}}{dt} = \mathbf{M} \times \left( -\gamma \mathbf{H}_{eff} + \frac{\alpha}{M_S} \frac{d\mathbf{M}}{dt} + \frac{\eta}{M_S} \frac{d^2\mathbf{M}}{dt^2} \right) \qquad (1)$$

and is called the inertial Landau-Lifshitz-Gilbert (ILLG) equation, where $\gamma = 2.2 \times 10^5 \text{ rad} \cdot \text{m} \cdot \text{A}^{-1} \cdot \text{s}^{-1}$ is the gyromagnetic ratio, $\mathbf{M}$ is the magnetization, $M_S$ is the saturation magnetization, $\mathbf{H}_{eff}$ is the effective magnetic field, $\alpha$ is the Gilbert precession damping, and $\eta$ is the inertial parameter [2,4,6,7].

First, we study the natural oscillations of the magnetization in the time domain to determine their trajectories, and the characteristic frequencies of precession and nutation modified by exchange coupling. In our previous work [32], we obtained analytical solutions for the magnetization trajectories from the ILLG equation in the nondamping case, $\alpha = 0$, without any approximations. Here we extend that investigation to focus on the dynamics of damped magnetization, using an approximation based on the linearization procedure. Next, we derive the dynamic susceptibility tensor in the presence of coupling from the ILLG equation considering forced oscillations. Finally, by substituting the susceptibility tensor into Maxwell's equations, we obtain the dispersion relations, which demonstrate hybridizations of precession-nutation spin waves with electromagnetic modes in ferromagnets. The similar hybridization processes of ordinary spin waves were studied in planar multiferroic structures [33]. The resulting dispersion relations describe waves propagating in an arbitrary direction relative to the applied magnetic field. We show that inertia of magnetization yields nutation spin waves, while the usual spin waves are now subject to the nutational motion.

## II. MAGNETIZATION DYNAMICS IN THE PRESENCE OF EXCHANGE INTERACTION AND INERTIA

The rigorous solution of Eq. (1) is a rather complicated task, which can be solved in very limited cases. For example, an analytical solution for the undamped deterministic motion of magnetization was given in Ref. [32]. The inclusion of damping or exchange interaction makes the analytical solution extremely difficult. Fortunately, approximate methods are available for solving the problem in the general case. These methods are widely used to analyze the solution of an ordinary LLG equation [22]. Here we apply the method of successive approximations to linearize the ILLG, Eq. (1).

In the presence of exchange interaction, the effective magnetic field $\mathbf{H}_{eff}(\mathbf{r},t)$ and the magnetization $\mathbf{M}(\mathbf{r},t)$ can be expressed as

$$\mathbf{H}_{eff}(\mathbf{r},t) = \mathbf{H}_0 + \mathbf{h}(\mathbf{r},t), \tag{2}$$

$$\mathbf{M}(\mathbf{r},t) = \mathbf{M}_0 + \mathbf{m}(\mathbf{r},t), \tag{3}$$

where $|\mathbf{H}_0| \gg |\mathbf{h}|$, $|\mathbf{M}_0| \gg |\mathbf{m}|$, $\mathbf{H}_0 = \mathbf{H}_{ext} + \mathbf{H}_a$ incorporates a strong uniform external field $\mathbf{H}_{ext} = H_{ext}\mathbf{e}_z$ directed along the $Z$ axis and an internal field due to the internal anisotropy potential $\mathbf{H}_a$ (the $\mathbf{H}_a$ can be neglected in a strong uniform external field), and $\mathbf{M}_0 = M_0\mathbf{e}_z$ (again in a strong uniform external field directed along the $Z$ axis the magnetization $\mathbf{M}(\mathbf{r},t)$ of a sample is almost saturated in the direction of the field $\mathbf{H}_{ext} = H_{ext}\mathbf{e}_z$, namely $M_0 \approx M_S$). On substituting Eqs. (2) and (3) into Eq. (1) and linearizing the resulting equation, we obtain

$$\frac{d\mathbf{m}}{dt} = -\gamma \mathbf{m} \times \mathbf{H}_0 + \mathbf{M}_0 \times \left( -\gamma \mathbf{h} + \frac{\alpha}{M_S}\frac{d\mathbf{m}}{dt} + \frac{\eta}{M_S}\frac{d^2\mathbf{m}}{dt^2} \right). \tag{4}$$

Since we investigate natural oscillations without external excitation, the varying magnetic field is equal to the exchange field, viz.,

$$\mathbf{h}(\mathbf{r},t) = \mathbf{h}_{ex}(\mathbf{r},t) = \lambda \nabla^2 \mathbf{m}(\mathbf{r},t), \tag{5}$$

where $\lambda$ is the exchange spin-wave stiffness. With the harmonic ansatz, the varying component of the magnetization is written as

$$\mathbf{m}(\mathbf{r},t) = \mathbf{m}(t)e^{-i\mathbf{r}\cdot\mathbf{k}}, \tag{6}$$

and, consequently $\mathbf{h}(\mathbf{r},t) = -\lambda k^2 \mathbf{m}(t)e^{-i\mathbf{r}\cdot\mathbf{k}}$, where $k=|\mathbf{k}|$. Using Eqs. (5) and (6), we can rewrite Eq. (4) as

$$\frac{d\mathbf{m}}{dt} = -\left[ (\omega_H + \omega_M \lambda k^2)\mathbf{m} + \alpha \frac{d\mathbf{m}}{dt} + \eta \frac{d^2\mathbf{m}}{dt^2} \right] \times \mathbf{e}_Z, \tag{7}$$

where $\omega_H = \gamma H_0$ and $\omega_M = \gamma M_0$. Unlike the ordinary LLG equation, the linearized Eq. (1) is now a second-order differential equation. By projecting the vector Eq. (7) onto the laboratory (Cartesian) axes, one obtains the following equations:

$$\frac{dm_X}{dt} = -\left(\omega_H + \omega_M \lambda k^2\right) m_Y - \alpha \frac{dm_Y}{dt} - \eta \frac{d^2 m_Y}{dt^2}, \tag{8}$$

$$\frac{dm_Y}{dt} = \left(\omega_H + \omega_M \lambda k^2\right) m_X + \alpha \frac{dm_X}{dt} + \eta \frac{d^2 m_X}{dt^2}, \tag{9}$$

$$\frac{dm_Z}{dt} = 0. \tag{10}$$

Introducing the circular variables,

$$m_\pm = m_X \pm i m_Y, \tag{11}$$

we derive the equation of motion,

$$\eta \frac{d^2 m_\pm}{dt^2} + (\alpha \pm i) \frac{dm_\pm}{dt} + \left(\omega_H + \omega_M \lambda k^2\right) m_\pm = 0, \tag{12}$$

which together with Eq. (10) yields the following solutions:

$$m_+(t) = A_- e^{i\omega_- t} + A_+ e^{-i\omega_+ t}, \quad m_-(t) = m_+^*(t), \tag{13}$$

$$m_Z(t) = m_Z(0) = 0, \tag{14}$$

where the asterisk denotes complex conjugation,

$$\omega_\pm = \frac{\pm(1-i\alpha) + \sqrt{(1-i\alpha)^2 + 4\eta(\omega_H + \omega_M \lambda k^2)}}{2\eta}, \tag{15}$$

$$A_\pm = \frac{1}{2}\left(m_\pm(0) \pm \frac{2i\eta m'_\pm(0) - (1-i\alpha) m_\pm(0)}{\sqrt{(1-i\alpha)^2 + 4\eta\left(\omega_H + \omega_M \lambda k^2\right)}}\right), \tag{16}$$

$$m'_\pm(0) = \frac{dm_\pm}{dt}(0). \tag{17}$$

If $k = 0$, the frequency of ferromagnetic resonance is given by $\omega_-$, whereas nutation resonance corresponds to $\omega_+$. Note that in the limiting inertia-free case $\eta \to 0$, the amplitudes are given by

$$\lim_{\eta \to 0} A_+ = 0, \quad \lim_{\eta \to 0} A_- = m_+(0). \tag{18}$$

This means that nutation disappears if inertia is neglected.

The trajectories of the end of the magnetization vector are shown in Fig. 1. In the nondamping case, the trajectories are star-shaped (Figs. 1(a) and 1(c)). In the damping case, the trajectories converge to the center (Figs. 1(b), (d) and (f)). With the approximation $m_Z = 0$, the trajectories lie in a plane, and the condition of magnetization length conservation, $|\mathbf{M}| = M_S = \text{const}$, is broken.

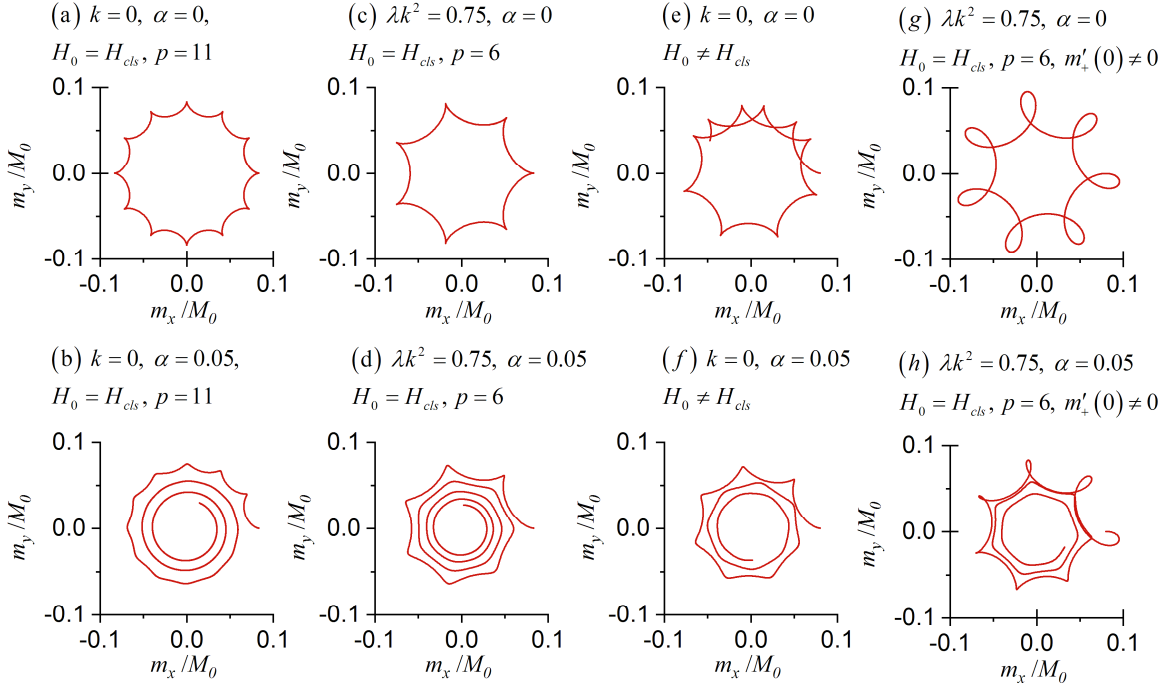

FIG. 1. Magnetization trajectories for $\mu_0 H_0 = 0.84$ T, $m'_+(0) = 0$ (a) and (b), $\mu_0 H_0 = 0.92$ T, $m'_+(0) = 0$ (c) and (d), $\mu_0 H_0 = 0.92$ T, $m'_+(0) = 3.82 \times 10^{16}$ A m$^{-1}$ s$^{-1}$ (g) and (h). The calculation was performed for $\mu_0 M_0 = 1.2$ T, $m_+(0) = 0.08 M_0$, $\eta = 0.75$ ps. Here $\mu_0 = 4\pi \cdot 10^{-7}$ (H/m) is the magnetic permeability of free space.

For the rigorous solution the trajectories are loop shaped in three-dimensional space [32]. The loop shaped trajectories in two-dimensional space can be obtained if the initial velocity of the magnetization vector is relatively high (Figs. 1(g), (h)). One can compare this velocity with the typical values of the first precession term of the ILLG equation. The magnitude of the spikes decreases as the inertial parameter $\eta$ decreases; they disappear at $\eta = 0$; therefore, the precessional trajectory can be observed in the limiting case (see, for example, Refs. [16,23]).

If a precession period $T_{pr}$ is equal to an integer number $p$ of nutation periods $T_{nu}$,

$$T_{pr} = pT_{nu}, \quad p \in \mathbb{N}, \tag{19}$$

then the trajectory is closed. By substituting $T_{pr} = 2\pi/\omega_-$ and $T_{nu} = 2\pi/\omega_+$ into Eq. (19), we obtain the magnetic field $H_{cls}$ satisfying such trajectories:

$$H_{cls} = \frac{1}{\gamma\eta}\frac{p}{(p-1)^2} - M_0 \lambda k^2. \tag{20}$$

The number of nutation spikes is proportional to $p$ and decreases as the wave number $k$ increases at the constant magnetic field (compare Figs. 1(a) and 1(c)). If the condition, Eq. (19), is not fulfilled, the trajectories are not closed (Fig. 1(e) and 1(f)). One can see that the magnetization trajectories, and hence the waveform, depend on the wave number and the effective magnetic field. The waveform can be visualized as a sequence of oscillating magnetic moments with phase shifts [13]. Note that the closed-loop condition means a rational ratio of the precession and nutation harmonics, which gains significance in the nonlinear wave dynamics.

### III. MAGNETIC SUSCEPTIBILITY MODIFIED BY EXCHANGE COUPLING

Equation (15) yields the characteristic frequencies of the magnetization dynamics. Now we apply an AC external magnetic field $\mathbf{h}_{app}(t) = \mathbf{h}e^{i\omega t}$ with small amplitude $\mathbf{h}$ to the ferromagnetic sample. If the AC field frequency $\omega$ coincides with the characteristic frequency of magnetization motion, resonance takes place in the spectrum of the susceptibility of the ferromagnetic sample. Next, we derive expressions for the susceptibility tensor $\hat{\chi}(\omega)$ introduced by the formula $\mathbf{m} = \hat{\chi}\mathbf{h}$ and used in the dispersion relations below. There are different methods for studying the susceptibility spectrum, namely, (i) direct Fourier transform [9] and (ii) the use of the harmonic ansatz. Since both methods give the same result in the case of linear differential equations, which we study here, the harmonic ansatz is employed. In accordance with Eq. (13), we may seek the time dependent component of the magnetization as $\mathbf{m}(t) = \mathbf{m}e^{i\omega t}$. The total time-varying field is now

$$\mathbf{h}(\mathbf{r},t) = \left(\mathbf{h}_0 - \lambda k^2 \mathbf{m}\right) e^{i\omega t - i\mathbf{k}\cdot\mathbf{r}}. \tag{21}$$

This field corresponds to a plane wave. Thus, vector Eq. (4) may be transformed into a system of scalar equation by analogy with Eqs. (8) and (9):

$$i\omega m_X + \left(\omega_H + \omega_M \lambda k^2 + i\omega\alpha - \omega^2 \eta\right) m_Y = \gamma M_S h_Y, \tag{22}$$

$$-i\omega m_Y + \left(\omega_H + \omega_M \lambda k^2 + i\omega\alpha - \omega^2 \eta\right) m_X = \gamma M_S h_X. \tag{23}$$

By rewriting these equations in vector form one derives the antisymmetric susceptibility tensor,

$$\begin{pmatrix} m_X \\ m_Y \\ m_Z \end{pmatrix} = \begin{pmatrix} \chi & i\chi_a & 0 \\ -i\chi_a & \chi & 0 \\ 0 & 0 & 0 \end{pmatrix} \begin{pmatrix} h_X \\ h_Y \\ h_Z \end{pmatrix}, \tag{24}$$

where

$$\chi = \frac{\omega_M (\omega_H + \omega_M \lambda k^2 - \eta\omega^2 + i\omega\alpha)}{(\omega_H + \omega_M \lambda k^2 - \eta\omega^2 + i\omega\alpha)^2 - \omega^2}, \tag{25}$$

$$\chi_a = \frac{\omega_M \omega}{(\omega_H + \omega_M \lambda k^2 - \eta\omega^2 + i\omega\alpha)^2 - \omega^2}. \tag{26}$$

As we noted before, the susceptibility tensor components are of resonant structure with the resonance frequencies $\omega_{res}$ calculated from the equation

$$\eta \omega_{res}^2 + (\pm 1 - i\alpha)\omega_{res} - \omega_H - \omega_M \lambda k^2 = 0, \tag{27}$$

namely,

$$\omega_{res} = \frac{\pm 1 + i\alpha \pm \sqrt{(1 \pm i\alpha)^2 + 4\eta(\omega_H + \omega_M \lambda k^2)}}{2\eta}. \tag{28}$$

In the nondamping case ($\alpha = 0$) at $k \to 0$, this expression yields the ferromagnetic and nutation resonance frequencies;

$$\omega_{FMR} = \frac{1}{2\eta}\left(\sqrt{1 + 4\eta(\omega_H + \omega_M \lambda k^2)} - 1\right) \approx \gamma H_0, \tag{29}$$

$$\omega_{NR} = \frac{1}{2\eta}\left(\sqrt{1 + 4\eta(\omega_H + \omega_M \lambda k^2)} + 1\right) \approx \frac{1}{\eta}. \tag{30}$$

The dispersion relation is based on the dependence of the susceptibility on the wave number. In order to clarify the physics of this dependence, we diagonalize the susceptibility tensor by using the circular variables $m_\pm(t)$ defined by Eq. (11), which obey

$$\begin{pmatrix} m_+ \\ m_- \\ m_z \end{pmatrix} = \begin{pmatrix} \chi_+ & 0 & 0 \\ 0 & \chi_- & 0 \\ 0 & 0 & 0 \end{pmatrix} \begin{pmatrix} h_+ \\ h_- \\ h_z \end{pmatrix}, \tag{31}$$

where $h_\pm = h_x \pm i h_y$ and

$$\chi_\pm = \chi \pm \chi_a = \frac{\omega_M}{\omega_H + \omega_M \lambda k^2 - \eta \omega^2 + i\omega\alpha \mp \omega}. \tag{32}$$

The susceptibility tensor is diagonal and $m_\pm = \chi_\pm h_\pm$. In the positive frequency range, $\chi_+$ exhibits ferromagnetic resonance with right-hand side rotation of magnetization (precession), whereas $\chi_-$ shows nutation resonance with opposite rotation. The resonance frequencies are determined by Eq. (15). The influence of the wave number on the susceptibility is demonstrated in Fig. 2 by separating the dispersive and dissipative parts of the susceptibility with $\chi_\pm = \chi'_\pm - i\chi''_\pm$, viz.,

$$\chi'_\pm = \frac{\omega_M(\omega_H + \omega_M \lambda k^2 - \eta\omega^2 \mp \omega)}{(\omega_H + \omega_M \lambda k^2 - \eta\omega^2 \mp \omega)^2 + \omega^2\alpha^2}, \tag{33}$$

$$\chi''_\pm = \frac{\omega_M \omega \alpha}{(\omega_H + \omega_M \lambda k^2 - \eta\omega^2 \mp \omega)^2 + \omega^2\alpha^2}. \tag{34}$$

The dispersive $\chi'_\pm$ and the dissipative $\chi''_\pm$ parts of the susceptibility for right-hand side (a) and left-hand side (b) rotations are shown in Figs. 2 and 3. The maximum of the curve $\chi''_\pm$ decreases as the wave number increases. One can see in Fig. 3 that the resonance frequency of the dissipative part of the susceptibility $\chi''_\pm$ exhibits a quadratic dependence on the wave number, namely, $\omega_{res} \approx \omega_H + \omega_M \lambda k^2$ for $\chi''_+$ and $\omega_{res} \approx \eta^{-1} + \omega_M \lambda k^2$ for $\chi''_-$. Moreover, the width of the curve $\chi''_\pm$ measured at a definite level decreases as the wavenumber increases. Thus, precessional $\chi_+$ and nutational $\chi_-$ susceptibilities demonstrate qualitatively similar behavior, but the latter has the substantially weaker resonance in the terahertz frequency range.

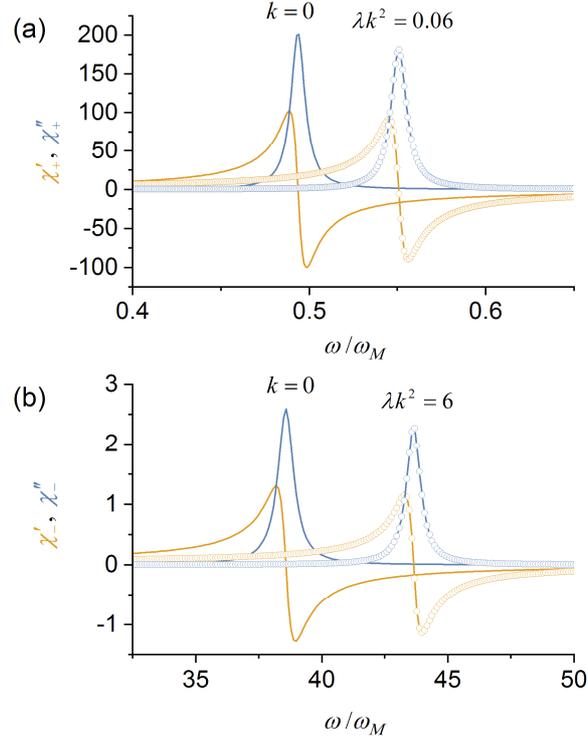

FIG. 2. The dispersive ($\chi'_\pm$ blue lines) and the dissipative ($\chi''_\pm$ orange lines) parts of susceptibility vs. $\omega/\omega_M$ for right-hand side (a) and left-hand side (b) rotations. The calculation was performed for $\mu_0 M_0 = 0.2$ T, $H_0 = 0.5 M_0$, $\eta = 0.75$ ps and $\alpha = 0.01$. The solid lines show susceptibility at $k = 0$, the lines with circles demonstrate susceptibility at $\lambda k^2 > 0$.

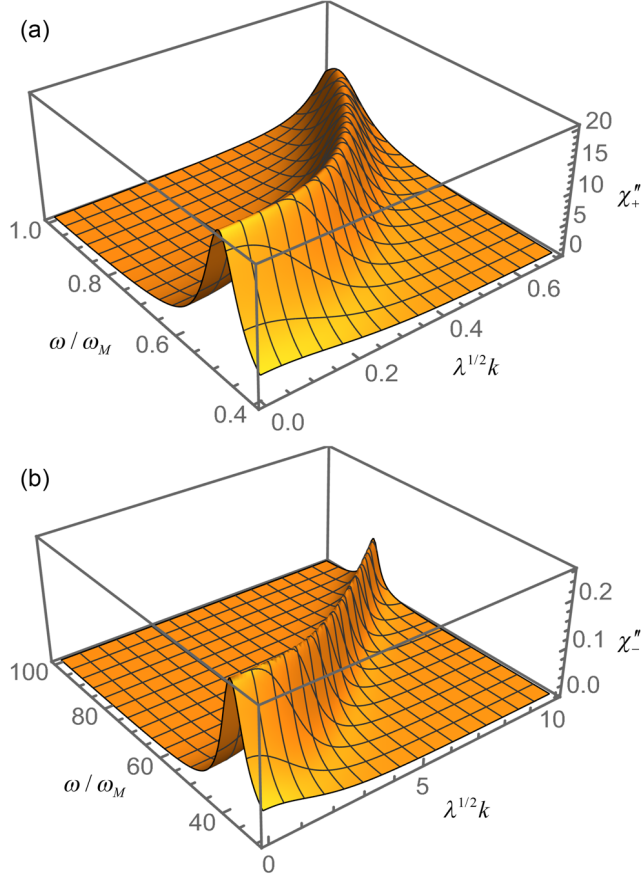

FIG. 3. The dissipative parts of susceptibilities $\chi''_+$ (a) and $\chi''_-$ (b) vs. frequency and wave number for $\mu_0 M_0 = 0.2$ T, $H_0 = 0.5 M_0$, $\eta = 0.75$ ps, and $\alpha = 0.1$.

## IV. DISPERSION RELATIONS

The dependence of the resonance frequency on the wave number, Eq. (28), is still not a dispersion relation, because such a relation should obey Maxwell's equations in ferromagnets. The derivation is briefly presented in the Appendix. The resultant equation for waves propagating in an arbitrary direction in a nonconductive magnetically anisotropic and electrically isotropic ferromagnet is [16]

$$n^4 \mu - n^2 \left[ (\mu-1) n_z^2 + \left( \mu^2 - \mu_a^2 + \mu \right) \varepsilon_r \right] + \varepsilon_r \left[ \left( \mu^2 - \mu_a^2 \right)\left( n_z^2 + \varepsilon_r \right) - \mu n_z^2 \right] = 0, \qquad (35)$$

where $\mathbf{n}$ is the dimensionless wave vector given by $\mathbf{n} = c\mathbf{k}/\omega$, $n = |\mathbf{n}|$, while $\mu$, $\mu_a$ are components of the permeability tensor $\hat{\mu}_r$ Eq. (A3), and $\varepsilon_r$ is the relative permittivity of the ferromagnet. The dispersion relation, Eq. (35), is used in the analysis of spin waves in ferromagnets with the dynamics of magnetization obeying the ordinary LLG equation. The difference in the present consideration is that the susceptibility tensor now takes into account the inertia of magnetization (see Eqs. (25) and (26) for the susceptibility tensor components).

## A. Waves propagating perpendicular to the applied magnetic field

In the general case, Eq. (35) can be solved numerically or by using the method for solving a quartic equation. Here, for simplicity, let us focus on two particular configurations: (A) waves propagating perpendicular to the uniform magnetic field ($n_z = 0$), and (B) waves propagating in the direction of $\mathbf{H}_0 \parallel \mathbf{e}_Z$ ($n = n_z$, $n_x = n_y = 0$). For the former (A) case, the general dispersion relation can be converted to

$$n^4 \mu - n^2 \varepsilon_r \left( \mu^2 - \mu_a^2 + \mu \right) + \varepsilon_r^2 \left( \mu^2 - \mu_a^2 \right) = 0, \tag{36}$$

which gives two roots

$$n^2 = \varepsilon_r \tag{37}$$

and

$$n^2 = \frac{(\mu^2 - \mu_a^2)\varepsilon_r}{\mu} = \frac{(1+\chi_+)(1+\chi_-)\varepsilon_r}{1+\chi}. \tag{38}$$

The first root applies to the electromagnetic waves caused by the dielectric properties of ferromagnets. The second root yields a set of dispersion branches, which can be considered with normalized variables: $\Omega = \omega/\omega_M$, $\Omega_H = \omega_H/\omega_M$, $\eta' = \eta\omega_M$, and $\lambda' = \lambda k_M^2$, where $k_M = \sqrt{\varepsilon_r}\omega_M/c$, $c = 1/\sqrt{\varepsilon_0\mu_0}$ is the speed of light, and $\varepsilon_0 \approx 10^{-9}/36\pi$ (F/m) is the electric permittivity of free space. From Eq. (38), one obtains the following expression for the normalized wave number $k' = k/k_M$:

$$k'^2 - \Omega^2 \frac{\left(\eta'\Omega^2 - i\alpha\Omega - \Omega_H - \lambda'k'^2 - 1\right)^2 - \Omega^2}{\left(\eta'\Omega^2 - i\alpha\Omega - \Omega_H - \lambda'k'^2\right)^2 - \left(\eta'\Omega^2 - i\alpha\Omega - \Omega_H - \lambda'k'^2\right) - \Omega^2} = 0, \tag{39}$$

which is the bicubic equation resulting in a few dispersion branches. Employing the notations

$$\begin{aligned}
a &= \lambda'^2, \quad b = \lambda'\left[2i\alpha\Omega - \Omega^2(2\eta' + \lambda') + 2\Omega_H + 1\right], \\
c &= -\Omega^2 + \Omega\left[-i\alpha + (\eta' + 2\lambda')\Omega\right]\left(-1 - i\alpha\Omega + \eta'\Omega^2\right) \\
&\quad + \Omega_H + 2i\alpha\Omega\Omega_H - 2(\eta' + \lambda')\Omega^2\Omega_H + \Omega_H^2, \\
d &= -\Omega^2\left[-1 + \Omega(-1 - i\alpha + \eta'\Omega) - \Omega_H\right]\left[-1 + \Omega(1 - i\alpha + \eta'\Omega) - \Omega_H\right],
\end{aligned} \tag{40}$$

$$\Delta_0 = b^2 - 3ac, \quad \Delta_1 = 2b^3 - 9abc + 27a^2d,$$

$$C = \left(\frac{\Delta_1 + \sqrt{\Delta_1^2 - 4\Delta_0^3}}{2}\right)^{1/3}, \quad \xi = \frac{-1+\sqrt{-3}}{2}, \tag{41}$$

we write the dispersion relations in Table I.

TABLE I. The dispersion equations for the waves propagating perpendicularly to the applied magnetic field

| Type of waves in the inertia case | Type of waves in the non-inertia case | Equation |
|---|---|---|
| Nutation electromagnetic (NEL) | vanishing | $k'_1 = \pm\sqrt{-\dfrac{1}{3a}\left(b + C + \dfrac{\Delta_0}{C}\right)}$, |
| precession-nutation magnetostatic (PNM) | precession magnetostatic (PM) | $k'_2 \pm \sqrt{-\dfrac{1}{3a}\left(b + \xi C + \dfrac{\Delta_0}{\xi C}\right)}$, |
| hybrid nutation (HN) | precession electromagnetic (PEL) | $k'_3 = \pm\sqrt{-\dfrac{1}{3a}\left(b + \xi^2 C + \dfrac{\Delta_0}{\xi^2 C}\right)}$, |

In the noninertia limit, one observes the ordinary spin waves hybridized with electromagnetic waves due to phase synchronism in the vicinity of $k' = 1$, $\Omega = 1$ for the calculation parameters (Fig. 4). The electromagnetic hybridization yields two dispersion branches [16], which we called here precession magnetostatic (PM) and precession electromagnetic (PEL) branches in order to distinguish these waves from the nutation ones. If one increases the wave number and moves away from the hybridization area, then the dispersion of the PEL waves asymptotically tends to the curve of electromagnetic waves, whereas the dispersion branch of the PM waves merges with the spin-wave curve:

$$\Omega^2 = \left(\Omega_H + \lambda' k'^2\right)\left(\Omega_H + \lambda' k'^2 + \sin^2\theta_k\right). \tag{42}$$

Inertia causes an additional dispersion curve starting at $1/\eta$ or $\Omega = 30$ for the ferromagnet under consideration (not shown in Fig. 4). This curve intersects the PEL branch at $k' = 30$ ($k = 140$ cm$^{-1}$) providing an extra hybridization (Fig. 4(b)), which eventually results in electromagnetic nutation waves described by nutation electromagnetic (NEL) and hybrid nutation (HN) branches. One can describe the behavior of the waves corresponding to the HN branch in the following manner. The dispersion branch of this wave starts at the frequency $\sqrt{\Omega_H(\Omega_H + 1)}$ or $\sqrt{\omega_H(\omega_H + \omega_M)}$ in non-normalized units, which is typical for magnetostatic spin waves in ferromagnets, then the HN branch exhibits electromagnetic behavior after the first hybridization (Fig. 4(c) and (d)); the second hybridization transforms these waves into the nutational ones, which are subject to exchange interaction at $\lambda k^2 > 1$. In addition, PM dispersion is redshifted by inertia to the dispersion of the precession-nutation magnetostatic (PNW) waves possessing relatively low frequencies (Fig. 4(e)).

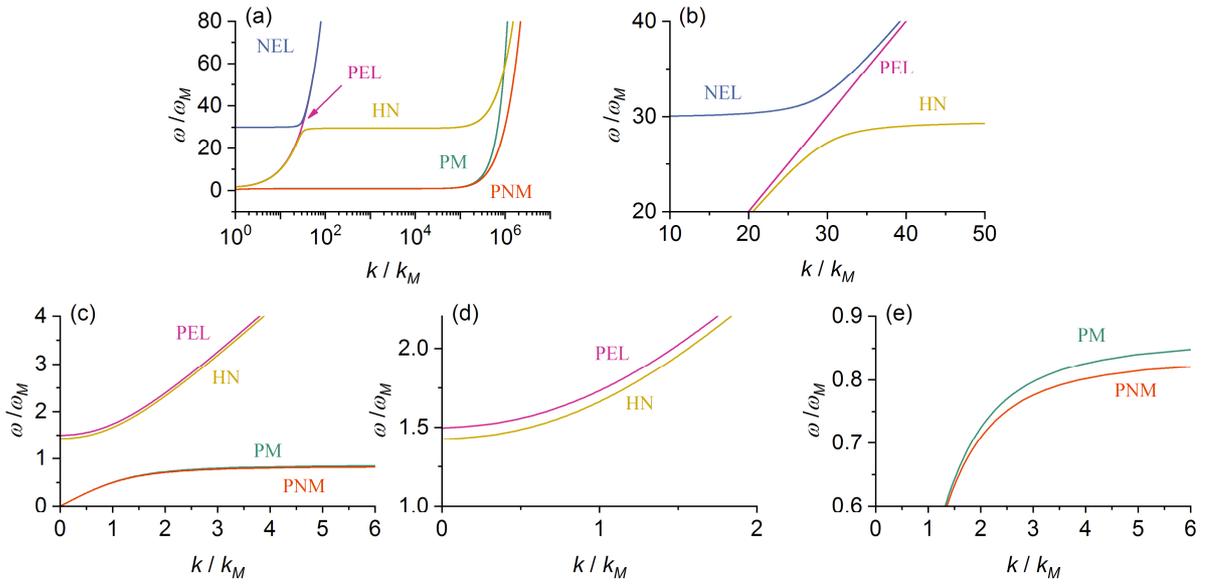

FIG. 4. Dispersion relations ($\omega/\omega_M$ vs $k_2/k_M$) for propagation perpendicular to the applied field ($\mathbf{k} \perp \mathbf{H}_0$) shown in different scales. The notation of curves: precession magnetostatic (PM, green), precession electromagnetic (PEL, magenta), precession-nutation magnetostatic (PNM, red), nutation electromagnetic (NEL, blue) and hybrid nutation (HN, yellow). The calculation is performed for $\mu_0 M_0 = 0.2$ T, $\omega_H/\omega_M = 0.5$, $\lambda = 3\cdot 10^{-16}$ m$^2$, $\varepsilon_r = 1.55$, $\alpha = 0$ and for $\eta = 1$ ps in the inertia case, $\eta = 0$ in non-inertia case.

The magnetic field of the waves can be calculated by substituting the dispersion relations into Eq. (A7). For the first root given by $n^2 = \varepsilon_r$ (Eq. (37)), the wave does not interact with the medium (since $\mu_\parallel = 1$; see Eq. (A3)) and propagates as in a nonmagnetic material. The substitution yields $\mathbf{h} = h_0 \mathbf{e}_z$ with an arbitrary value of $h_0$; hence we have

$$\mathbf{h}(\mathbf{r},t) = \mathbf{h}e^{i(\omega t - k_0 y)} = h_0 \mathbf{e}_z e^{i(\omega t - k_0 y)}. \tag{43}$$

The inertia of magnetization does not change the wave property in this case.

The substitution of the second root (Eq. (38)) into Eq. (A7) with $n = n_y$ and $n_x = 0$ leads to

$$\hat{A} = \varepsilon_r \begin{pmatrix} -\chi_a^2(1+\chi)^{-1} & -i\chi_a & 0 \\ i\chi_a & -1-\chi & 0 \\ 0 & 0 & \chi - \chi_a^2(1+\chi)^{-1} \end{pmatrix}. \tag{44}$$

The solution of Eq. (A7) with $\hat{A}$ determined by Eq. (44) is given by

$$\mathbf{h}(\mathbf{r},t) = \begin{pmatrix} 1 \\ i\chi_a(1+\chi)^{-1}\big|_{k=k_i} \\ 0 \end{pmatrix} h_0 e^{i(\omega t - k_i y)}, \tag{45}$$

where $k_i$ is the root of Eq. (39). Thus, the PM, PEL, PNM, NEL, and HN waves propagating perpendicular to the applied field $\mathbf{H}_0$ possess an elliptically polarized magnetic field $\mathbf{h}$. Since the components of the susceptibility tensor are modified by the inertia, the magnitude of the polarization depends on the inertia parameters.

### B. Waves propagating parallel to the applied magnetic field

For the waves propagating parallel to the direction of the applied magnetic fields, we set $n = n_z$ and $n_x = n_y = 0$. Therefore, the general dispersion relation becomes

$$n^4 - 2n^2 \mu \varepsilon_r + \left( \mu^2 - \mu_a^2 \right) \varepsilon_r^2 = 0 \tag{46}$$

and gives two roots

$$n^2 = \left( \mu + \mu_a \right)\varepsilon_r = \left( 1 + \chi_+ \right)\varepsilon_r, \quad n^2 = \left( \mu - \mu_a \right)\varepsilon_r = \left( 1 + \chi_- \right)\varepsilon_r \tag{47}$$

which in turn results in four dispersion equations:

$$k'^2_{+\pm} = \frac{(\eta' + \lambda')\Omega^2 - (i\alpha - 1)\Omega - \Omega_H}{2\lambda'}$$
$$\pm \frac{\sqrt{\left((\eta' + \lambda')\Omega^2 - (i\alpha - 1)\Omega - \Omega_H\right)^2 - 4\lambda'\left(\eta'\Omega^2 - (i\alpha - 1)\Omega - \Omega_H - 1\right)\Omega^2}}{2\lambda'}, \tag{48}$$

$$k'^2_{-\pm} = \frac{(\eta' + \lambda')\Omega^2 - \Omega(i\alpha + 1) - \Omega_H}{2\lambda'}$$
$$\pm \frac{\sqrt{\left((\eta' + \lambda')\Omega^2 - \Omega(i\alpha + 1) - \Omega_H\right)^2 - 4\lambda'\left(\eta'\Omega^2 - (i\alpha + 1)\Omega - \Omega_H - 1\right)\Omega^2}}{2\lambda'} \tag{49}$$

describing different types of waves (Table II). From Fig. 5, one can see that changing the propagation angle barely affects the dependence of the frequency on the wave number of PEL, NEL, PM and PNM waves, except the frequency shift due to the $\sin^2 \theta_k$ factor. In contrast, the HN branch undergoes an abrupt transformation in the vicinity of the first electromagnetic hybridization: now it starts from $\omega = 0$. In addition, a precession-nutation electromagnetic (PNEL) dispersion branch emerges, which is the PEL branch redshifted by inertia.

In order to find the polarization of the waves in this parallel case, we substitute two roots given by Eq. (47) into Eq. (A7) and obtain

$$\hat{A}_\pm = \varepsilon_r \begin{pmatrix} \pm \chi_a & -i\chi_a & 0 \\ i\chi_a & \pm \chi_a & 0 \\ 0 & 0 & -1 \end{pmatrix}. \tag{50}$$

The solution of Eq. (A7) with $\hat{A}_\pm$ determined by Eq. (50) is given by

$$\mathbf{h}(\mathbf{r},t) = \begin{pmatrix} 1 \\ \mp i \\ 0 \end{pmatrix} h_0 e^{i(\omega t - k_{\pm} y)}. \tag{51}$$

Consequently, the magnetic fields of waves propagating parallel to the applied field $\mathbf{H}_0$ become circularly polarized.

TABLE II. The correspondence between the dispersion equations and type of waves

| Type of waves in the inertia case | Type of waves in the non-inertia case | Equation |
|---|---|---|
| precession-nutation magnetostatic (PNM) | precession magnetostatic (PM) | $k'_{++}$, Eq. (48) |
| precession-nutation electromagnetic (PNEL) | precession electromagnetic (PEL) | $k'_{+-}$, Eq. (48) |
| hybrid nutation (HN) | electromagnetic (EL) | $k'_{-+}$, Eq. (49) |
| Nutation electromagnetic (NEL) | vanishing | $k'_{--}$, Eq. (49) |

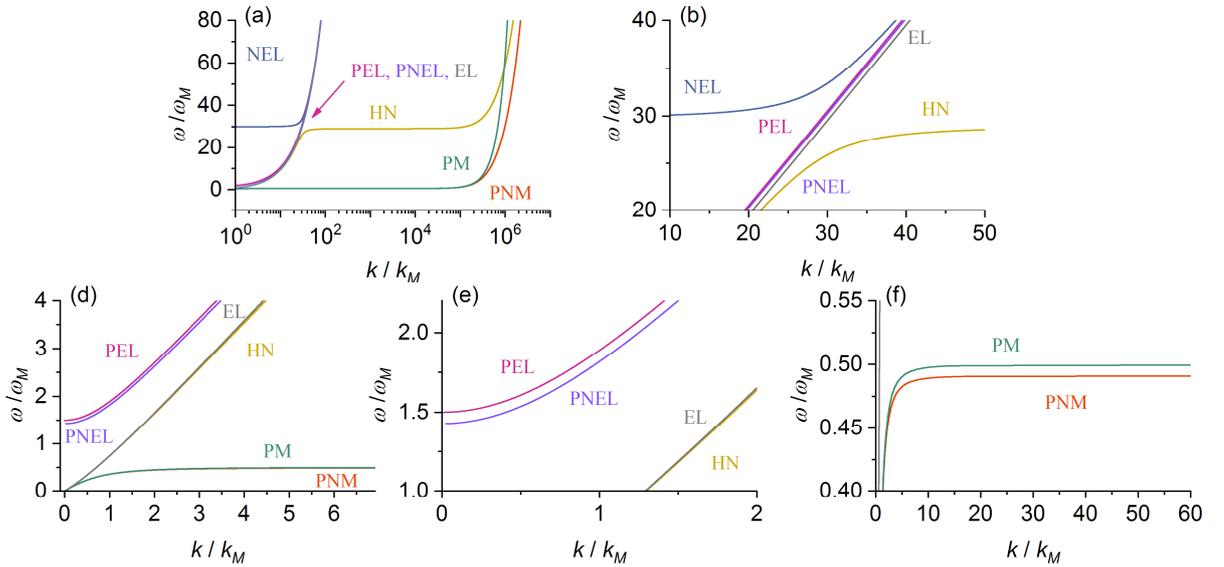

FIG. 5. Dispersion relations ($\omega/\omega_M$ vs $k_2/k_M$) of waves propagating parallel to the applied field ($\mathbf{k} \parallel \mathbf{H}_0$) shown in different scales. The notation of curves: precession magnetostatic (PM, green), precession electromagnetic (PEL, magenta), precession-nutation magnetostatic (PNM, red), nutation electromagnetic (NEL, blue), precession-nutation electromagnetic (PNEL, violet) electromagnetic (EL, grey) and hybrid nutation (HN, yellow). The calculation is performed for $\mu_0 M_0 = 0.2$ T, $\omega_H/\omega_M = 0.5$, $\lambda = 3 \cdot 10^{-16}$ m$^2$, $\varepsilon_r = 1.55$, $\alpha = 0$ and for $\eta = 1$ ps in the inertia case, $\eta = 0$ in the non-inertia case.

# V. CONCLUSION

We have considered the propagation of plane waves in ferromagnets, when the magnetic material is close to saturation and the contribution of the internal anisotropic potential of the ferromagnetic material can be neglected. We have found that the trajectories of magnetization are star-shaped with the approximation $m_z = 0$. The number of nutation spikes decreases as the wave number increases. Moreover, we demonstrated that the trajectories can be either closed or open, depending on the wave number and the effective magnetic field. Note that the presented trajectories are deterministic. If the thermal fluctuations are included, the trajectories should be averaged over the initial conditions by using the Boltzmann distribution in the equilibrium state [34]. We envisage that inertia can cause the chaotic behavior of magnetization and we intend to address this question in the future.

We have shown that the inertia of magnetization causes the additional hybridization of electromagnetic waves and nutation spin waves in ferromagnets, hybrid nutation spin waves emerge, and the redshift of frequencies of precession spin waves is initiated, which transforms to precession-nutation spin waves. Note that these effects depend on the direction of wave propagation relative to the applied magnetic field. The magnetic field of the investigated spin waves is elliptically polarized in the case when the waves propagate perpendicular to the applied field, and the waves become circularly polarized in the parallel configuration.

It is of interest to generalize the results obtained by including magnetocrystalline and shape anisotropies as well as the finite size of ferromagnets, as was performed for the ferromagnetic and nutation resonance frequencies [35, 36]. The further complication of the wave spectrum can be predicted in this task. The presented methods can be further developed to treat tensor electrical properties, nonlinearity of nutation spin waves, and conductivity of ferromagnets.

We thank P.-M. Déjardin for a critical reading of the manuscript and useful suggestions.

# APPENDIX: DERIVATION OF THE GENERAL FORM OF THE DISPERSION RELATION

The electromagnetic field in a nonconductive magnetically anisotropic and electrically isotropic medium is governed by Maxwell's equations which have the form [16]

$$\nabla \times \mathbf{H} = \varepsilon_0 \varepsilon_r \frac{\partial \mathbf{E}}{\partial t}, \quad \nabla \times \mathbf{E} = -\mu_0 \hat{\mu}_r \frac{\partial \mathbf{H}}{\partial t}, \tag{A1}$$

$$\nabla \cdot \mathbf{E} = 0, \quad \nabla \cdot \mathbf{B} = 0. \tag{A2}$$

Here, $\varepsilon_r$ is the relative permittivity of the medium, $\varepsilon_0 \approx 10^{-9}/36\pi$ (F/m) is the electric permittivity of free space, and $\mu_0 = 4\pi \cdot 10^{-7}$ (H/m) is the magnetic permeability of free space. The magnetic permeability tensor associated with the derived susceptibility Eqs. (25) and (26) is written as

$$\hat{\mu}_r = \begin{pmatrix} \mu & i\mu_a & 0 \\ -i\mu_a & \mu & 0 \\ 0 & 0 & \mu_{\|} \end{pmatrix}, \quad (A3)$$

where $\mu = 1+\chi$, $\mu_a = \chi_a$ and $\mu_{\|} = 1$ (saturation, $M_0 = M_S$).

Taking the curl ($\nabla \times$) of the first equation of Eqs. (A1) and excluding $\mathbf{E}$ by using the second equation of Eqs. (A1), we obtain [18] ($\varepsilon_0 \mu_0 = c^{-2}$)

$$\frac{\varepsilon_r \hat{\mu}_r}{c^2} \frac{\partial^2 \mathbf{H}}{\partial t^2} + \nabla \times \nabla \times \mathbf{H} = 0 \quad (A4)$$

If one expresses the external magnetic field and magnetization as previously (Eqs. (2)-(3)) and employs the plane-wave ansatz,

$$\mathbf{M}(\mathbf{r},t) = \mathbf{M}_0 + \mathbf{m} e^{i(\omega t - \mathbf{r}\cdot\mathbf{k})}, \quad (A5)$$

$$\mathbf{H}(\mathbf{r},t) = \mathbf{H}_0 + \mathbf{h} e^{i(\omega t - \mathbf{r}\cdot\mathbf{k})}, \quad (A6)$$

Eq. (A4) transforms into

$$\frac{\omega^2}{c^2} \hat{A}\mathbf{h}(\mathbf{r},t) = 0, \quad (A7)$$

where the matrix $\hat{A}$ is given by

$$\hat{A} = \begin{pmatrix} n^2 - n_x^2 - \mu\varepsilon_r & -n_x n_y - i\mu_a \varepsilon_r & -n_x n_z \\ -n_x n_y + i\mu_a \varepsilon_r & n^2 - n_y^2 - \mu\varepsilon_r & -n_y n_z \\ -n_x n_z & -n_y n_z & n^2 - n_z^2 - \varepsilon_r \end{pmatrix} \quad (A8)$$

and $n_x, n_y, n_z$ are the components of the dimensionless wave vector $\mathbf{n}$ defined as

$$\mathbf{n} = \frac{c}{\omega} \mathbf{k}. \quad (A9)$$

The dispersion relation obtained from the equation $\det \hat{A} = 0$ is given by Eq. (35).